# Asymmetries of Secondary Cosmic Muons with High Statistics and Low Systematics

T. F. Lin, B. Fields*, J. Gress, and J. Poirier
*Physics Department, 225 NSH, University of Notre Dame, Notre Dame, IN 46556, USA*

## Abstract

Project GRAND measures the asymmetries of secondary cosmic muons by calculating the differences between the muons from the east and the muons from the west for complete solar days. A similar north-south difference is also presented. These muons are identified in single track data taken with the 64 stations of Project GRAND; the data cover a two year span with a total muon sample of 30 billion muons.

## 1 Introduction:

While galactic cosmic rays are isotropic to a high degree, departures from isotropy do exist. The unique design of Project GRAND's detectors is particularly well-suited to measure these anisotropies with high precision and low systematics.

For example, the Compton-Getting effect (CGE) (Compton & Getting, 1935; Culter & Groom, 1991) describes cosmic ray anisotropy of order $10^{-4}$ produced by the motion of the earth relative to a frame in which the cosmic rays are isotropic. At the equator, the induced anisotropy, in the direction $\theta$ relative to the earth's orbital motion, is $\Delta\phi(\theta)/\phi = (2+\gamma)(v_o/c)\cos(\theta)$, where $\gamma \cong 2.7$ is the cosmic ray spectral index, and $v_o$ = 29.7 km/s is the earth's orbital speed around the sun. Thus the amplitude of the effect is:

$$[\Delta\phi(\theta)/\phi]_{max} = \Delta\phi(0)/\phi = (2+\gamma)(v_o/c) = 4.7 \times 10^{-4}.$$

Thus GRAND is expected to have a contribution to its diurnal anisotropy of $4.7 \times 10^{-4}$ due to the CGE.

The difference between muons from the east and from the west is measured. The difference in E - W between noon and midnight (i.e. the double subtraction) allows determination of the CGE variations due to the earth's orbital motion around the sun. While studying the CGE, the search has been generalized for the possibility of other asymmetric effects which might appear in the data. By utilizing this double subtraction technique, almost all possible systematic effects are eliminated; statistics are large enough to get results with small statistical errors even after subtracting two nearly equal and large numbers.

## 2 Experimental Array:

Project GRAND, located at 220 m above sea level at 86° W and 42° N, detects cosmic ray secondaries at ground level by means of 64 tracking stations of proportional wire chambers (PWCs). Four x-y planes (Linsley et al., 1987; Poirier et al., 1990) arranged above each other measure each track to 0.26° absolute precision (average value in each of two orthogonal planes)(Gress et al., 1991). Each plane has an active area of 1.25 $m^2$ with 80 detection cells 10 mm high, 14 mm wide. A 50 mm steel plate above the bottom x-y PWC pair allows each track to be identified as a muon (or not); 4% of electron tracks are misidentified as muon tracks and 96% of muons are identified as muons. Since for single track data the muon tracks are four times the number of electron tracks, then 99% of the single tracks which pass the muon algorithm are real muons (Gress et al., 1990). The muon threshold energy is 0.1 GeV for vertical tracks, increasing as approximately $1/\cos(\theta)$ for $\theta$-angles inclined from vertical. The current rate for recording and identifying muons is 2000 Hz; in two years, 100 billion such muons have been collected with their angles measured and times recorded and reduced to solar and stellar angle coordinates; the analysis below is only for solar effects. After a smoothness test is imposed (described below), the number of muons in the final data sample is reduced to 30 billion.

# 3   Data Analysis:

For the past two years, Project GRAND has collected 289 data files with each data file covering more than one solar day. Interruptions can occur in the data due to various causes such as detector malfunction or a single station turned off for repair with the run continuing with the remaining stations. To ensure the data files are smooth through the whole solar day without inconsistencies from these interruptions, a smoothness test is conducted on all 289 data files. This smoothness test operates in the following manner:
The average muon count is calculated for each 48 half-hour interval, as well as the maximum and minimum values. The maximum deviation is calculated by taking the difference of the maximum and minimum which is then divided by the average count. To be included in the final data analysis, the maximum deviation for each data file had to be smaller than ± 13%. There were 198 solar days' data out of 289 data files which passed this smoothness test. This yielded a final data sample of 30 billion muons.

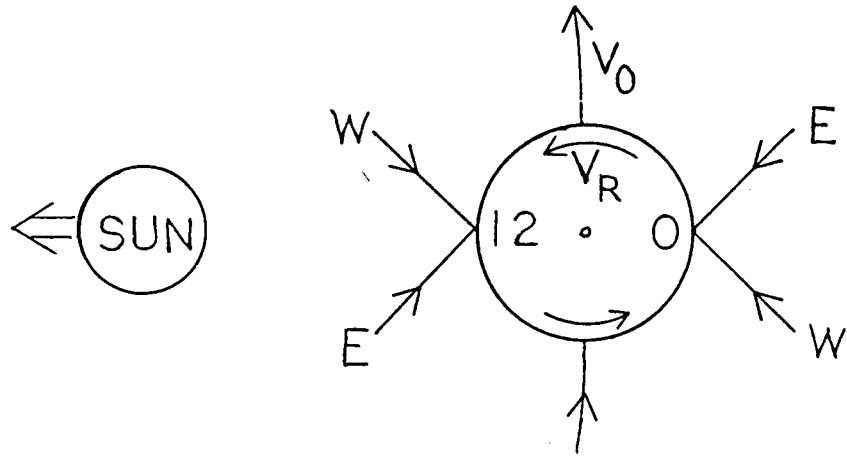

**Figure 1:** Earth's rotational speed $V_R$ about its axis, orbital speed $V_o$ about the sun, and local directon of cosmic rays from the east (E) or west (W). Viewed from the north pole; 0=midnight, 12=noon.

When Project GRAND faces toward the sun (at local *noon* time, see Fig. 1), muons from the west are coming from a direction toward the earth's motion about the sun while muons from the east come from a direction away from the earth's motion about the sun. This situation is reversed when Project GRAND faces away from the sun (midnight). Project GRAND studies $N_e$ (the number of cosmic rays arriving *from* the east) minus $N_w$ (*from* the west) termed E-W asymmetry and similarly, $N_n$ minus $N_s$ (north - southward) termed the N-S asymmetry, where each subtraction is divided by the average for that time interval. The east-west difference of muon count rates is computed for each 1/2-hour interval. By subtracting the E - W difference at noon from the same difference at midnight, a measurement of the CGE is obtained for the velocity of the earth about the sun; almost all other systematic effects are canceled in these subtractions. A large loss of statistical accuracy in this double subtraction is quite tolerable due to the large number of raw counts; the loss of statistical accuracy is counterbalanced by the confidence that almost all systematic errors have been eliminated. Using this double subtraction technique, this large data sample allows small asymmetries to be measured with good statistics while virtually eliminating all systematic errors.

The duration of the data sample allows an average over two years so that while studying *solar* effects, the *sidereal* effects can be averaged out. Conversely, the duration of the data sample allows an average over two complete sidereal years so that while studying *sidereal* effects, the *solar* effects can be averaged out. The analysis presented here is only for the *solar* effects with the *sidereal* effects averaged out.

The E - W asymmetry (muon angles *from* the west minus those *from* the east divided by the total number of muons) is observed (Fig. 2A) in 1/2-hour intervals and fit with the function:

$$ASYM = A + B\cos[15(x+C)] + D\cos[15(2x+E)]$$

where x, C, and D are in solar hours from midnight; the factor of 15 expresses the argument of the cosine in degrees from 0 to 360°. The north-south difference (N-S) is plotted in Fig. 2B and fit with the same function. All fit parameters are shown in Table 1 with their errors.

Table 1: Coefficients of the function ASYM which were fit to the data

|   | E-W asymmetry | N-S asymmetry |
| --- | --- | --- |
| A | $-(33.1 \pm 0.1) \times 10^{-4}$ | $+(5.5 \pm 0.1) \times 10^{-4}$ |
| B | $-(1.7 \pm 0.1) \times 10^{-4}$ | $-(2.9 \pm 0.1) \times 10^{-4}$ |
| D | $+(2.1 \pm 0.1) \times 10^{-4}$ | $+(2.3 \pm 0.1) \times 10^{-4}$ |
| C | $(2.7 \pm 0.2)$ hour | $(4.9 \pm 0.1)$ hour |
| E | $(0.3 \pm 0.2)$ hour | $(4.2 \pm 0.2)$ hour |

The value for A in both the E-W and N-S asymmetries contains systematic errors which are not eliminated as they are for all other coefficients--for example, a dead wire on the top PWC on the east side of a station would cause a uniform deficiency in angles from the east in that station causing a systematic shift in "A". "A" for the E - W asymmetry is a measure of the effect of the earth's magnetic field on the positively charged primary cosmic rays; it retains possible systemmatic effects. This same defect would not cause *changes* as a function of solar time which are parameterized by the constants B,C,D,and E which are therefore independent of possible systemmatics. The term $\cos[15(x+C)]$ models effects which vary once per day while $\cos[15(2x+E)]$ models those effects which change twice per day. The CGE would contribute to the component "B" in the E-W asymmetry.

## 4 Conclusions:

The CGE studied here relates to the earth's motion about the sun. The maximum effect in ASYM is at noon and midnight. To compare the results in Table 1 with the predicted CGE, one must account for the angular acceptance of GRAND. This acceptance is a maximum at zenith angle and minimum in the direction of the CGE. If each track's projection is taken along the earth's orbital velocity and multiplied by GRAND's acceptance and the $\cos^2\phi$ muon angular distribution (where $\phi$ is the track's angle from local zenith), then ASYM is 9.4% sensitive to the CGE. Multiplying the theoretical CGE number by GRAND's detection efficiency (9.4%) yields an ASYM value of $-0.4 \times 10^{-4}$ in the component direction of the CGE. The fact that an amplitude of $B = -1.7 \times 10^{-4}$ is found suggests there are asymmetric effects over and above the CGE effect. This remaining asymmetry, $-1.3 \times 10^{-4}$, is therefore due to physics other than CGE. The ASYM value of $-2.9 \times 10^{-4}$ for N-S has no CGE component, so this value also is due to other physics.

A possible explanation for the once-per-day solar effects is the solar magnetic field; a possible explanation for the twice-per-day solar effect are the air tides with two high tides and two low tides per day due to the gradient in the solar gravitational field (with the effect of the moon averaged out).

This research is presently being funded through grants from the University of Notre Dame and private donations. The National Science Foundation participated in GRAND's construction.

## References and Footnotes

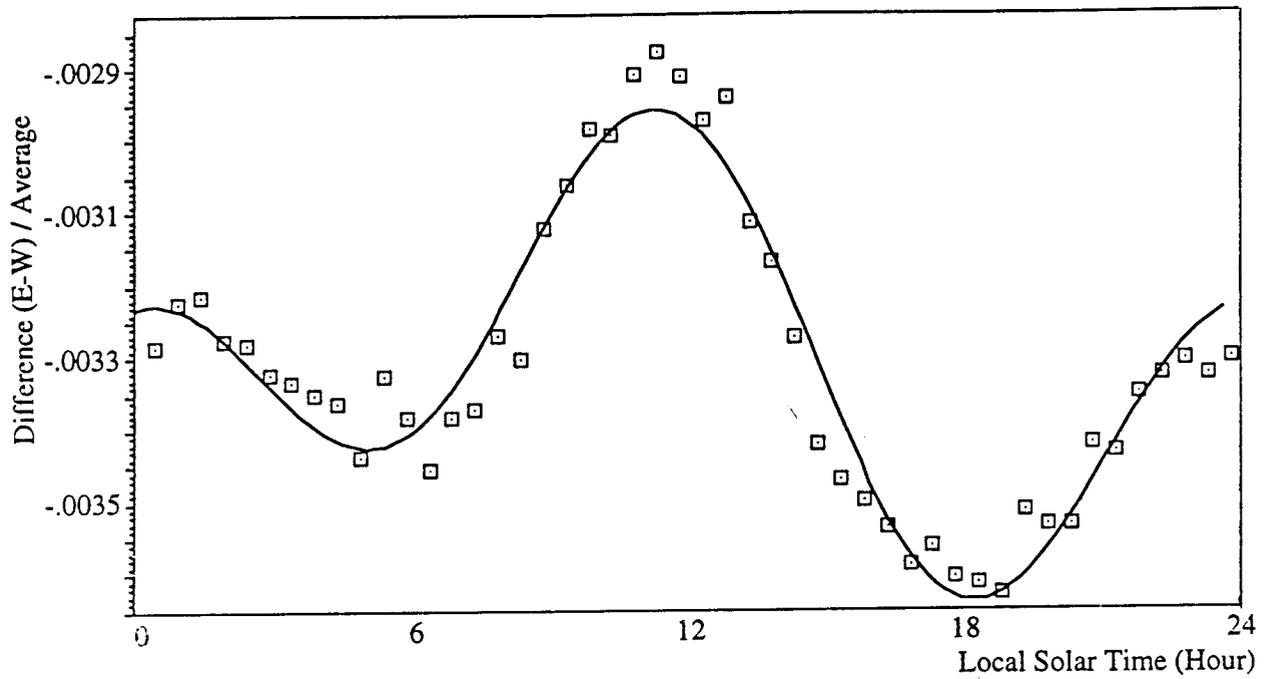

**Figure 2A**: Difference between muons from the east minus west divided by the average versus local solar time.

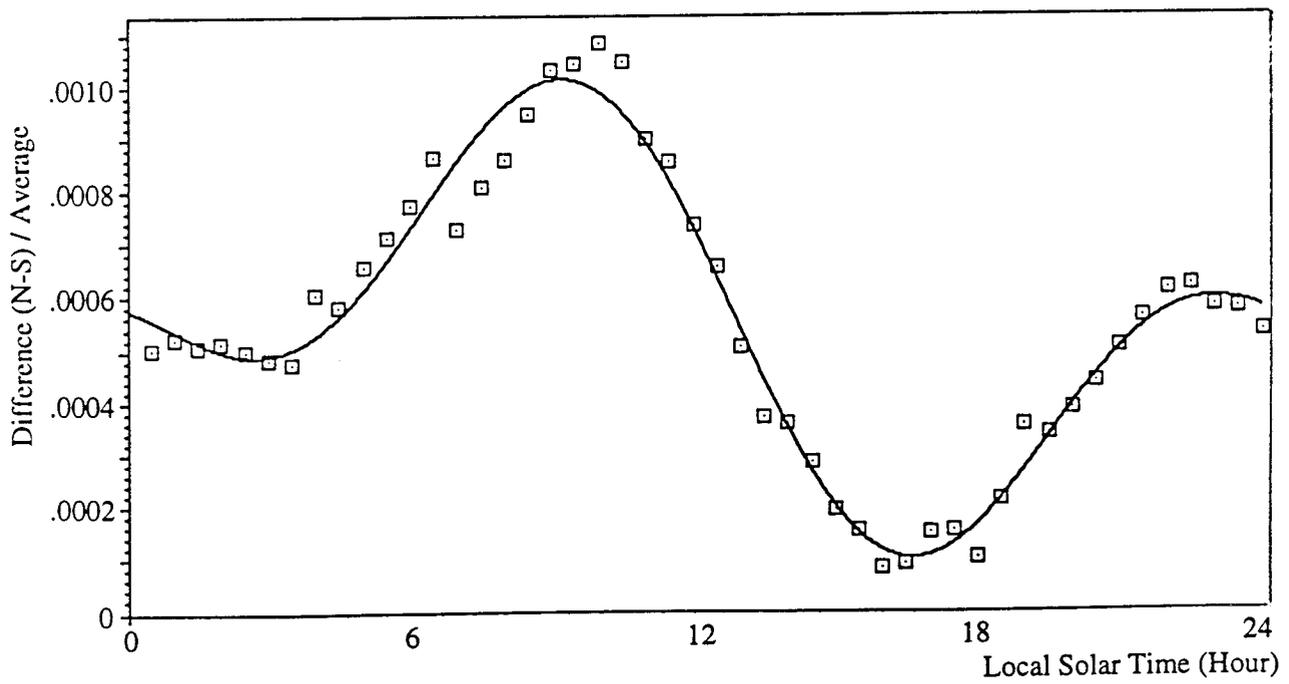

**Figure 2B**: Difference between muons from the north minus south divided by the average versus local solar time.